\documentclass[epj]{svjour}
\usepackage{graphicx}
\usepackage{amsfonts}
\usepackage{amsmath}
\usepackage{mathrsfs}

\newcommand{\antineu}  {$\bar \nu_\mathrm{e}$\ }

\begin{document}

\title{
Performances and stability of a 2.4 ton Gd organic liquid scintillator target for \antineu detection
}
\author{
I.R.Barabanov\inst{1} \and L.B.Bezrukov\inst{1} \and C.Cattadori\inst{2} \and N.A.Danilov\inst{1,3} \and A.Di Vacri\inst{4} \and Yu.S.Krilov\inst{1,3} \and L.Ioannucci\inst{4} \and E.A.Yanovich\inst{1} for the \textit{MetaLS} collaboration and\\
M.Aglietta\inst{5} \and A.Bonardi\inst{5} \and G.Bruno\inst{4} \and 
W.Fulgione\inst{5} \and
E.Kemp\inst{6} \and
A.S.Malguin\inst{1} \and A. Porta\inst{5} \and M.Selvi\inst{7}\\ 
for the \textit{LVD} Collaboration
}
\institute{
Institute for Nuclear Research RAS\\ 60th October Anniversary prospect 7a, Moscow 117312 - Russia \and INFN Milano Bicocca\\
Piazza della Scienza 3, 20126 Milano - Italy \and Institute of Physical Chemistry and Electrochemistry RAS, Leninsky Prospect 31,Moscow 119991- Russia \and INFN Laboratori Nazionali del Gran Sasso\\ S.S. 17 bis km 18.910
67010 Assergi L\rq Aquila - Italy \and INAF, Institute of Physics of Interplanetary Space\\ Corso Fiume 4, 10133 Torino - Italy \and Universidade Estadual de Campinas - UNICAMP\\ IFGW-DRCC 6165 CEP 13070-718 Campinas (SP) - Brasil \and INFN Bologna\\ Via Irnerio 46, 40126 Bologna - Italy
}
\date{Received: date / Revised version: date}

\abstract{
In this work we report the performances and the chemical and physical properties of a $2 \times 1.2$  ton organic liquid scintillator target doped with Gd up to\mbox{$\sim 0.1\%$}, and the results of a 2 year long stability survey. In particular we have monitored the amount of both Gd and primary fluor actually in solution, the optical and fluorescent properties of the Gd-doped liquid scintillator (GdLS) and its performances as a neutron detector, namely neutron capture efficiency and average capture time.  The experimental survey is ongoing, the target being continuously monitored. After two years from the doping time the performances of the Gd-doped liquid scintillator do not show any hint of degradation and instability; this conclusion comes both from the laboratory measurements and from the \textit{in-tank} measurements. This is the largest stable Gd-doped organic liquid scintillator target ever produced and continuously operated for a long period.
}
\maketitle

\section{Introduction} \label{intro}
Since the early phases of experimental neutrino physics, liquid scintillator targets loaded with a metal enhancing the neutron capture process, have been adopted to detect \antineu via inverse beta decay. However, the practical use of this technique has been hampered by the lack of stability of the chemical, optical and fluorescent properties of the loaded scintillator.
Low energy neutrino experiments as SN gravitational collapse, solar and reactor neutrino, typically last from few years (reactor experiments) to few year-decades (SN experiments), therefore the stability requirements of the target are compulsory.
Instability and degradation of Gd or Cd loaded scintillators have been reported by many authors, starting from the Hanford and Savannah River experiments \cite{Hanford,Savannah River,Ronzio}; then in 1965 Nezrick  and Reines reported instability of the Gd-doped scintillator caused by incompatibility with the container material \cite{Nezrick}.

A high efficiency 1.8 ton scintillation counter for neutron detection \cite{Bezrukov1} has been operated with Gd-octoate diluted in a mixture of hydrocarbons known as \emph{White Spirit}: this scintillator had a limited lifetime due to poor solubility of Gd-octoate in \emph{White Spirit} \cite{Olga1}.

In more recent years, the CHOOZ  \antineu experiment observed a chemical instability of the Gd-doped LS target ($\sim$ 5 tons of Norpar(50\%)+IPPB+hexanol doped at 0.1\% with Gd), probably due to the oxidation of the Gd salt, Gd(NO$_{3}$)$_{3}$ \cite{CHOOZ scint}, resulting into the degradation of the light transmittance. This forced a weekly measurement of the detector optical attenuation length that decayed exponentially with a lifetime of $\sim$ 240 days.
The transparency degradation caused a 35\% reduction in the photoelectron yield, from the beginning to the end of the experiment.

Finally, the fully blended Gd loaded liquid scintillator for the Palo Verde experiment \cite{PALO VERDE 2001}, contained in 200 l steel drums, turned out to be unstable during the transportation over ground via truck from Ohio to Arizona \cite{PALO VERDE scint}. The scintillator base was a mixture of 40\% Pseudocumene (1,2,4-Trimethyl-benzene) and 60\% mineral oil, doped up to 0.1\% in weight with Gd by the compound gadolinium 2-\-ethyl\-hexano\-ate. All the Gd-doped scintillator was totally replaced with a new one produced on-site adding a concentrated Gd-doped pseudocumene-based solution at the mineral oil . The experiment took data for about two years.

In years from 2000 to 2005 an intense R\&D activity on Yb, In, Gd loaded scintillators was started up and pursued by a joint INFN-INR group at LNGS, in the framework of LENS R\&D project, devoted to low energy solar neutrino physics \cite{Raghavan Yb,Rag In,LENS results}. This group investigated and understood the origins of the instabilities of metal loaded organic liquid scintillators, and developed methodics to avoid them \cite{Yb LS,In LS1,LENS scintillatore 1,LENS scintillatore 2}. In 2007 on the base of INFN-INR group the INFN MetaLS project has been started up.  

Finally in 2005 a practical procedure to produce large quantities (O($1$ t)) of stable Gd loaded scintillator was mature \cite{Gdsaltpaper}, and results of laboratory scale experiments on long term stability were reliable.

At the same time the LVD\footnote{LVD is a 1000 ton liquid scintillator detector \cite{LVD}, installed in the LNGS, mainly dedicated to the study of supernova neutrinos \cite{on-line}.} collaboration was studying the possibility to improve the detector performances \cite{opzione}; thus the two efforts were put together and it was decided to dope 2 modules (tanks) of the LVD detector \cite{Proceeding,tesiDivacri}.

The main advantage of Gd doping the LS of LVD is to improve the S/N ratio in the detection of the inverse beta decay \antineu interactions by lowering the neutron capture time of a factor of $\sim 10$ and increasing the energy of the emitted gamma rays ($\sim 8$ MeV instead of $2.2$ MeV).

The main goals of Gd doping 2 LVD counters are:
\begin{itemize}
	\item evaluate on ton scale the performances of the doped liquid scintillator; 
	\item monitor the stability when scintillator is kept in contact for few years with detector materials at not controlled environmental conditions (large temperature and humidity variability);
	\item study the feasibility of Gd-doping an existing large scale apparatus.
\end{itemize}

It is worthwhile to mention that in the recentest years an intense experimental activity \cite{DoubleChooz,Hahn,DayBay} is ongoing aiming to produce performant and stable GdLS targets, in the framework of new generation reactor neutrino experiments devoted to measure $sin^2\theta_{13}$, the last missing neutrino mixing angle.

\section{Properties of the liquid scintillator and the adopted Gd doping technology} \label{sec:1}
The unloaded liquid scintillator of the LVD experiment is a mixture of aliphatic and aromatic hydrocarbons (C$_{n}$H$ _{2n}$ with $\overline{n}=9.6$), also known as \emph{White Spirit} or \emph{Ragia Minerale}. There are two main kinds of LS: one (LS1) contains $\sim 16\%$ of aromatics and the other (LS2) $\sim 8\%$. They show slightly different characteristics, the higher percentage of aromatics causing a higher light yield \cite{tesiDivacri}. For both of them the light attenuation length at 425 nm, is $\Lambda \ge16$ m. 
A 15 kg batch of the organic Gd compound (Gd-2Methylvalerate) was prepared at LNGS chemical laboratory in spring 2005, following the recipe developed \cite{Gdsaltpaper}. Thanks to the simplicity and high reproducibility of the developed formulation \cite{tesiDivacri}, the whole 15 kg batch has been produced in one month using laboratory equipment. The adopted salt formulation was chosen to be highly soluble in organic solvent having low aromatic content.

Each LVD counter is a $1.5~m^3$ stainless steel tank containing 1.2 ton of LS. To dope such a large amount of LS handling a minimum aliquot of it, a master solution highly Gd-doped was first prepared and subsequently diluted in the tank: 31 l of LS were extracted from one counter, doped at a Gd concentration of 48 g/l level and then poured back in the counter. 
We chose to dope two counters: one with LS1 and the other with LS2. The Gd concentrations differ of about $15 \%$; concentration of stabilizing agent are different as well. The characteristics of the two doped counters are listed in the following:
\begin{enumerate}
\item T40, located in the Mounting Hall of the LNGS external laboratory (1000 m over the sea level), LS1 was doped at [Gd]= 1.05 g/l in May 2005. 
\item T3131, located in the underground laboratory (depth  3600 m w.e.), LS2 was doped at [Gd]=0.93 g/l in October 2005. 
\end{enumerate}

\section{The stability survey by laboratory measurements} \label{sec:2}
The 2 doped tanks have been periodically sampled, sucking some liquid ($\approx$ 0.5 l) from top and from bottom of the counter with a long clean inert pipe, to study eventual inhomogeneities of Gd concentration and/or optical transmittance, fluors concentrations, light yield, and survey these parameters.\\ 
LS from T40 was extracted on May and October 2005, January and  November 2006, January and June 2007, for a total of 6 extractions.\\
LS from T3131 was extracted on October 2005, February and November 2006, January and June 2007, for a total of 5 extractions.\\
In each extraction separate samples were taken from top (TOP) and bottom (BOTTOM) of tanks; moreover, to check whether some salt exits from solution, the February 2006 extraction of tank T3131 was performed before and after flushing the liquid scintillator for 8 hours by Argon gas, with the purpose to mix up the salt eventually deposited at the counter bottom.
Results of all the performed measurements are reported in the following sections.
\subsection{The optical properties} \label{subsec:2.1}
\begin{figure}[h] 
\begin{center}
\includegraphics[width=8cm]{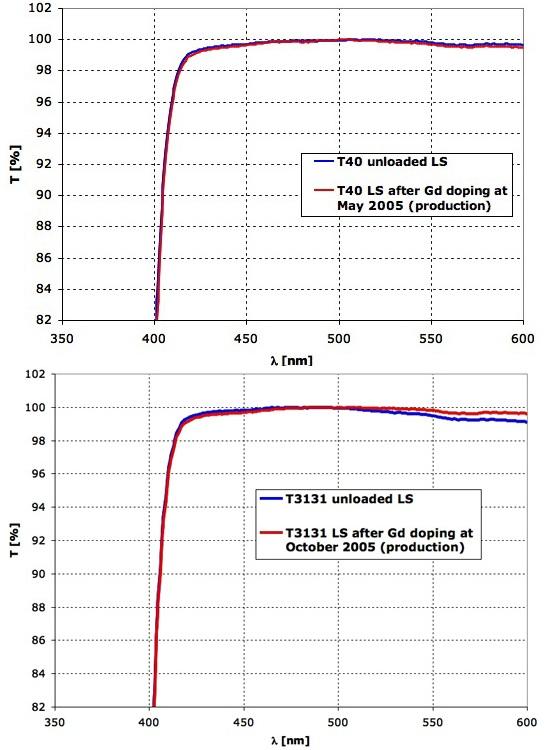}
\caption{Comparison of undoped (blue) and Gd-doped (red) T spectra of LS from tank T40 (top) and tank T3131 (bottom): Gd-doped sample have been extracted from the tanks 8 hours after doping. To mix up the 1.2 ton volume the tanks have been flushed by Ar.} \label{dop_vs_undop}
\end{center}
\end{figure}

\begin{figure}[h] 
\begin{center}
\includegraphics[width=7.5cm]{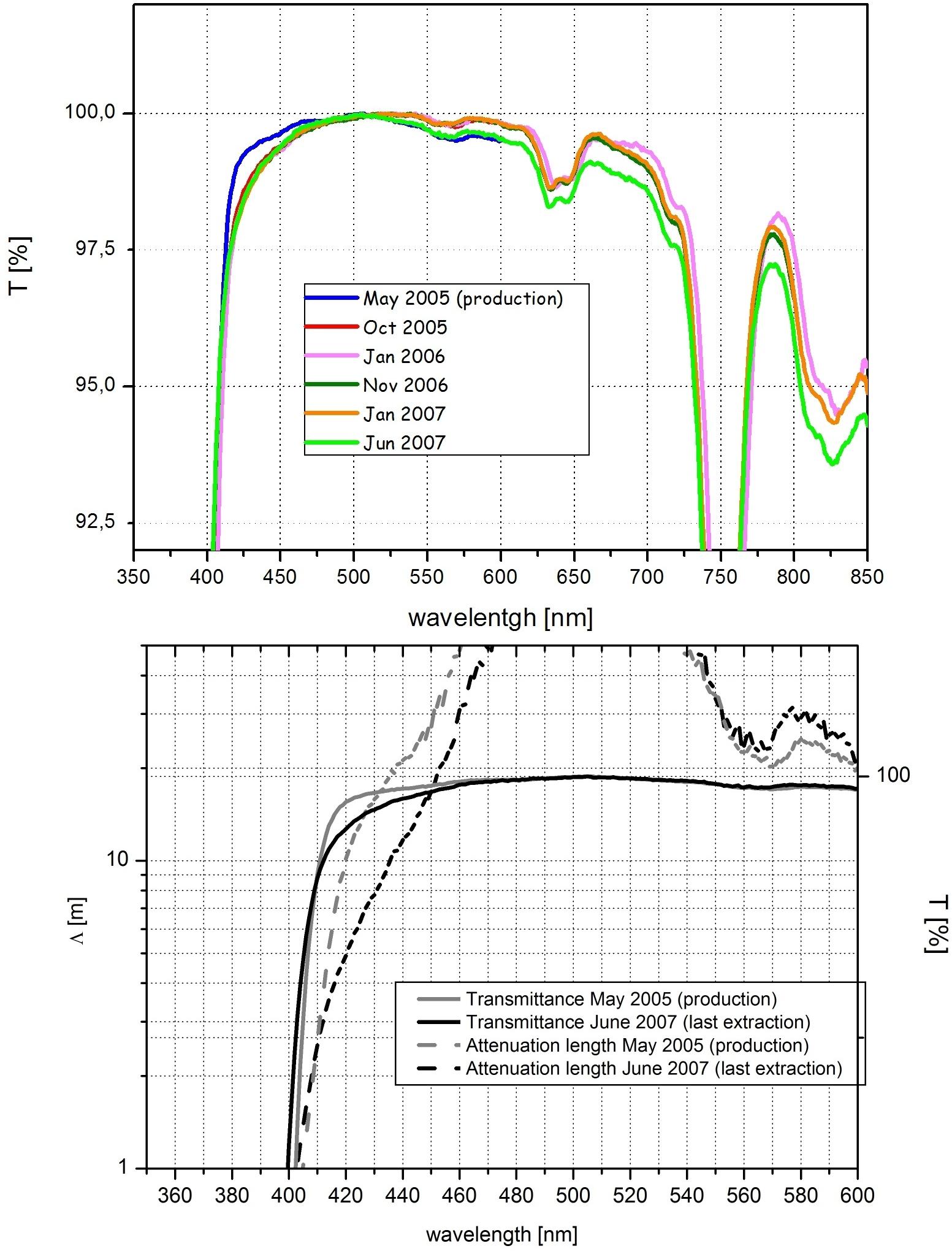}
\caption{Transmittance spectra (top) of Gd-doped LS samples extracted periodically from tank T40, and the derived light attenuation length (bottom). In the latter the attenuation length of first (may 2005) and last (june 2007) extractions are compared.} \label{T40_survey}
\end{center}
\end{figure}

\begin{figure}[htbp] 
\begin{center}
\includegraphics[width=7.2cm]{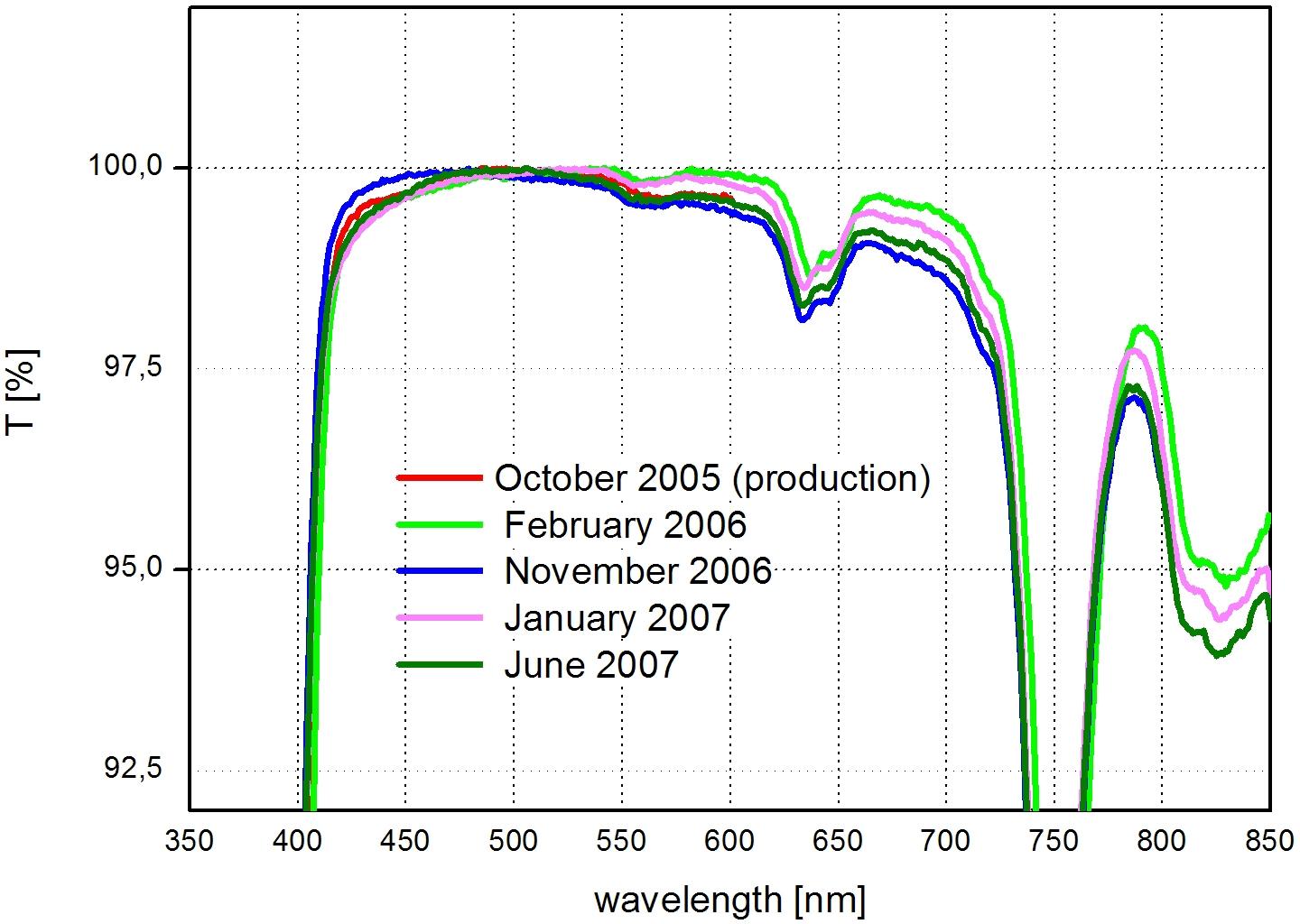}\\
\includegraphics[width=7cm]{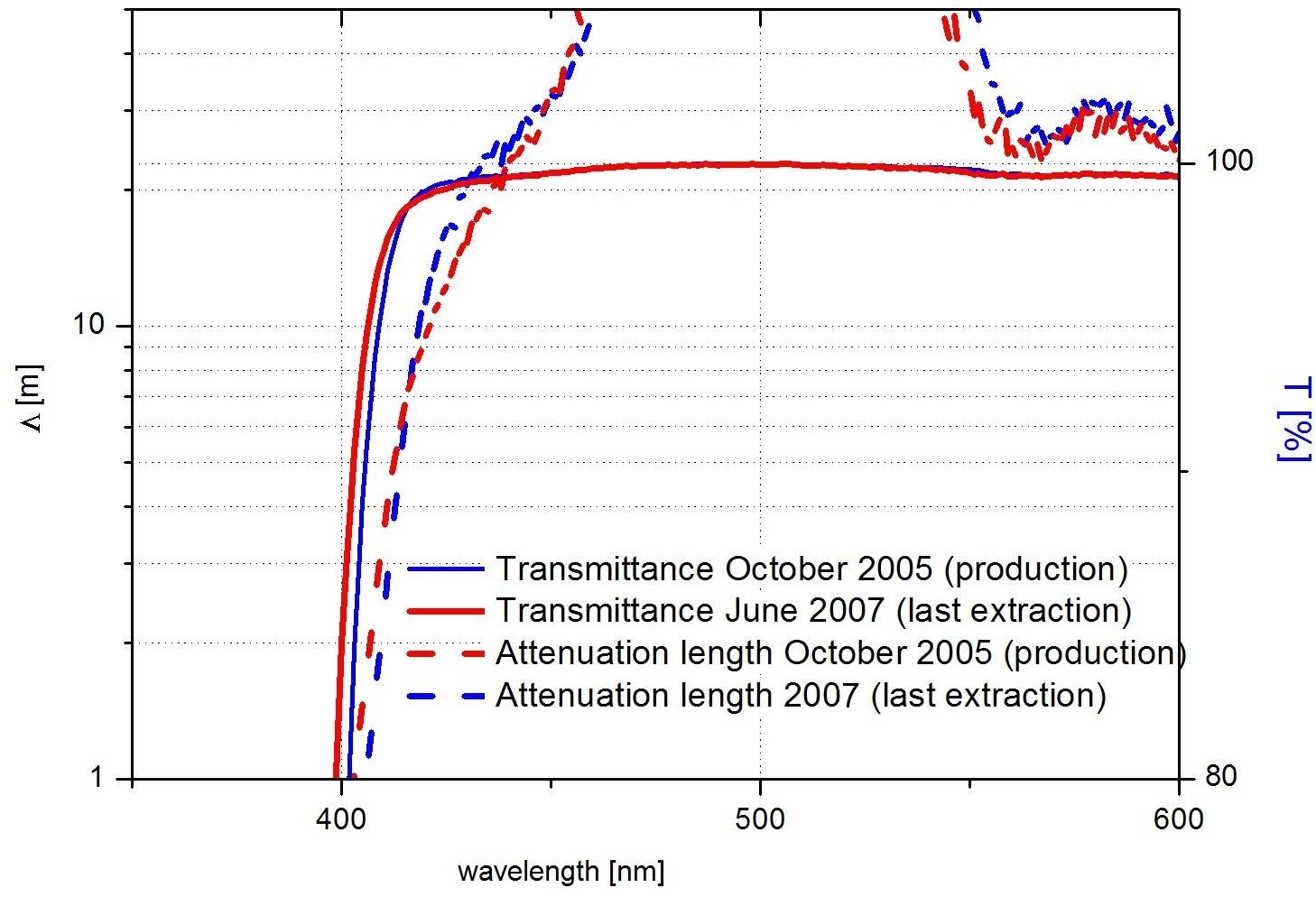}
\caption{Transmittance spectra (top) of Gd-doped LS samples extracted periodically from tank T3131, and the derived light attenuation length (bottom). The attenuation length is compared between first (october 2005) and last (june 2007) extraction.} \label{T3131_survey}
\end{center}
\end{figure}

The optical quality of the Gd-doped LS is evaluated measuring its light transmittance, by UV/VIS spectrophotometry, in the wavelength range from 350 to 800 nm. Measurements are performed in 10 cm optical path cuvettes with a Perkin Elmer Lambda 18 double beam spectrophotometer.\\
Light transmittance (T), defined as the ratio of the outcoming ($I$) to incoming ($I_0$) light beam intensity on a slice of media of $x$ cm optical path ( $T=\frac{I}{I_0}$), is related to the optical path $x$ [cm] and to the concentration $c$ [mole] of light-absorbing species present in the sample through the molar extintion coefficient $\epsilon$ by the Lambert-Beer law (eq.~\ref{Lambert-Beer}). $\epsilon$ is characteristic of the substance in analysis and depends on wavelength ($\lambda$); it is expressed in [mole$^{-1}\cdot$cm$^{-1}$]. The light absorbance is defined as $A=-\log T$. Therefore
\begin{equation}
A\left(\lambda\right)=-\log\left(T\right)=\left(\dfrac{\beta}{\ln 10}\right)c x=\epsilon\left(\lambda\right)c x\label{Lambert-Beer}
\end{equation}
The light attenuation length ($\Lambda$) defined as the distance at which the original light intensity $I_0$ is reduced of factor $\frac{1}{e}$, is related to A by the following law
\begin{equation}
\Lambda\left(\lambda\right)=\dfrac{x}{A\left(\lambda\right)\ln10}\label{latt}
\end{equation}
and can be easily determined once the transmittance spectrum is measured.
The systematic error of T spectrophotometric measurements can be reduced down to $\sim 1\%$ 
\\
Figure \ref{dop_vs_undop} shows for both tanks the T spectra of undoped scintillator compared to the Gd-doped sample extracted 8 hours after the doping time; just before the extraction the LS has been mixed by Ar flushing. The spectra overlap perfectly. The conclusion is that the developed Gd complex does not affect immediately the optical properties of the LS for $\lambda \geq 400$ nm.
Figure \ref{T40_survey} and fig. \ref{T3131_survey} show the measured transmittance spectra (left) and derived attenuation lengths (right) of samples extracted along 2 years from T40 and T3131.

It is important to point out that when measuring a low light-absorbing medium over an optical path length of only 10 cm, the values of $\Lambda$ deduced from T suffer of a large error, proportional to $(\ln{(1/T}))^{-1}$; for T $\sim 98 \%$ and $\Delta T/T$ =1\% the ($\Delta\Lambda/\Lambda$) is $50\%$. For this reason when $\Lambda$ is larger than 10 m, we use the latter value as a lower limit. When looking at fig. \ref{T40_survey}, one sees that Gd doping has a visible impact on LS1 in the wavelength range 410-450 nm, reducing T, in all the samples extracted in the  observation period, of $\sim 2\%$; in each spectrum the deduced $\Lambda$ at 425 nm is below the 10 m value, {\it{i.e.}} from 6 m to 7 m ($\pm 3$ m). This effect is observed in all samples. Nevertheless the main point for the purpose of this work is that no continuous degradation of T ($\Lambda$) is observed after a first assessment of the Gd complex in detector environment; from the T spectra collected from October 2005 till June 2007 there is no indication of a continuous degradation process. 

Based on the measurement of first sample extracted from tank T40, when tank T3131 was doped, a larger quantity of stabilizing agent was added to the master solution. For target T3131 (see fig. \ref{T3131_survey}), the Gd doping delayed impact on T is well below 1\%, and this indication comes from all the samples; the value of $\Lambda$ deduced from each of them is $\ge$ 10 m
for wavelengths $\geq$ 420 nm. 

\begin{figure}[htbp] 
\begin{center}
\includegraphics[width=6cm]{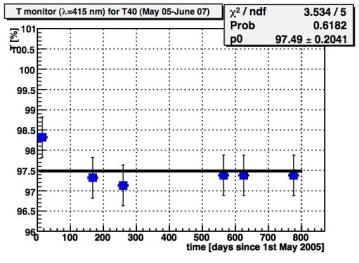}\\
\includegraphics[width=6cm]{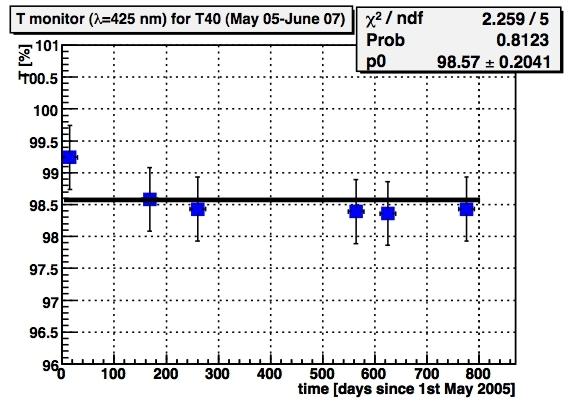}\\
\includegraphics[width=6cm]{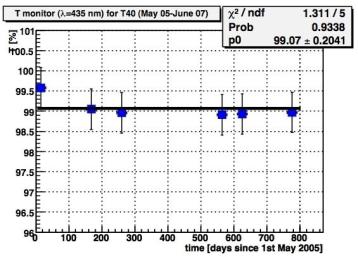}\\
\includegraphics[width=6cm]{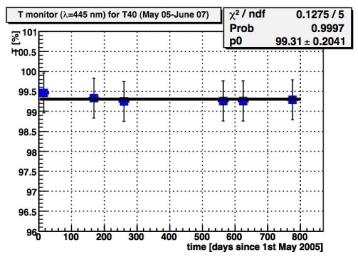}
\caption{ The value of the light transmittance measured for aliquots of Gd LS periodically extracted from tank T40 at wavelengths of 415 nm, 425 nm, 435 nm and 445 nm (from top to bottom) plotted as a function of time since Gd doping.} 
\label{fit_T_T40}
\end{center}
\end{figure}

\begin{figure}[htbp] 
\begin{center}
\includegraphics[width=6cm]{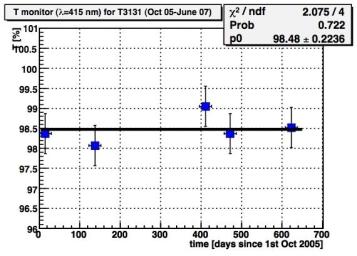}\\
\includegraphics[width=6cm]{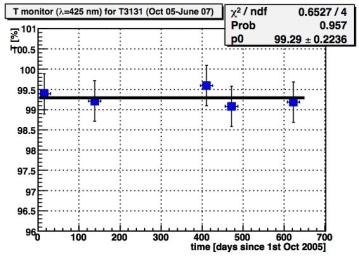}\\
\includegraphics[width=6cm]{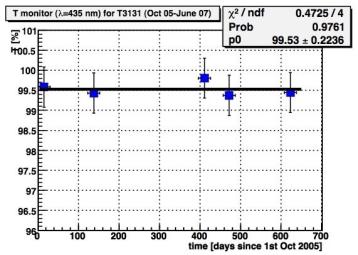}\\
\includegraphics[width=6cm]{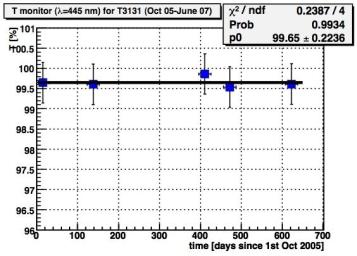}
\caption{The value of the light transmittance measured for aliquots of Gd LS periodically extracted from tank T3131 at wavelengths of 415 nm, 425 nm, 435 nm and  445 nm (from top to bottom) plotted as a function of time since Gd doping.} 
\label{fit_T_T3131}
\end{center}
\end{figure}

To quantitatively study the stability of the light transmittance in the relevant range of wavelengths, we plot the transmittance values at 415, 425, 435, 445 nm versus time, and fit the set of values in the hypothesis of constant behaviour. Figure \ref{fit_T_T40} and fig. \ref{fit_T_T3131} show the results respectively for tank T40 and tank T3131. The stability of T at all the wavelengths is confirmed by the fit results; at the reference wavelength of 425 nm the average value of T over the whole period is $\langle T\rangle$ = $98.6\pm 0.2$\% ($\Lambda = 7 \pm 1$ m) and $\langle T\rangle$ = $99.3\pm 0.2$ \% ($\Lambda = 14 \pm 4$ m), with a C.L. of 81\% and 96\% respectively for tank T40 and tank T3131. For tank T40, where a first interaction of Gd-complex in detector environment has been observed, excluding from the data set the T value of the sample extracted at doping time (therefore before interaction took place) the C.L. increases up to 99\%.   

T spectra of TOP and BOTTOM samples from both targets (spectra not reported here) do not show any appreciable difference in the whole wavelength range, therefore indicating an optical homogeneity.


\subsection{The Gd and primary fluor concentrations}\label{fluorconc}

The Gd concentration in the scintillator is determinated by complexometric EDTA titration after back-extraction, performing 3 independent titrations on each sample. We measured [Gd] = 1.05$\pm$0.02 g/l on each T40 sample, and [Gd] = 0.93$\pm$0.02 g/l on each T3131 sample. No difference has been observed between TOP and BOTTOM samples; therefore we can state that the Gd concentration in the tanks is homogeneous and stable since the doping time at 2\% level.

\begin{figure}[htbp] 
\begin{center}
\includegraphics[width=8cm]{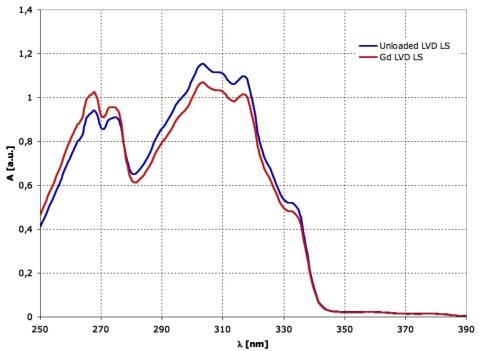}
\caption{Absorbance spectra of an undoped LS sample (blue line) and of a Gd-doped ([Gd]=0.1\%) one (red line), in the UV wavelength region. No extra absorption bands are introduced by the developed Gd complex.} 
\label{UVspectrum}
\end{center}
\end{figure}

The investigation of the impact of Gd doping on the fluors already diluted in the LVD LS has been perfomed by UV/VIS spectrophotometry and by gas-chromatography. First we have measured the Gd-complex molar extintion coefficient $\epsilon_{\rm{Gd-complex}}$ in the relevant wavelength range between 250 and 350 nm; it is $10^{-4}$ of the PPO molar extintion coefficient $\epsilon_{\rm{PPO}}$. Figure \ref{UVspectrum} shows that the developed Gd-complex does not introduce any new absorption band, that quenches the scintillation light, affecting the non-radiative energy transfer to primary fluor 
; the absorbance spectra of undoped and doped LS overlap almost perfectly. It can be noticed that unexpectedly the doped sample (for both tanks) has lower absorbance than the undoped one, in the range corresponding to PPO absorbance band (280-340 nm). This observation triggered the 
quantitative determination of PPO in the doped LSs, by both gas-chromathography (GC) and UV/VIS spectrophotometry. Table \ref{measfluorconc} reports the results: we observe that the introduction of the Gd complex clearly affects the free-fluor concentration of $\sim 30\%$ in tank T40, and of $\sim 10 \%$ in tank T3131. The effect is confirmed by the two independent techniques. Since no precipitate is visible in the extracted samples, the observed reduction of fluor concentration suggests that a Gd-complex - fluor interaction takes place, to make a new complex that is no more (or less) optically active; this will affect the quantitative determination by both techniques, as the calibration curves are obtained from free-fluor fixed concentration solutions. This effect has never been reported for Gd-doped LSs and needs further detailed investigation. The reduction of the free and active fluor concentration is consistent with the measured impact ($\sim 10\%$) of Gd doping on the light yield (see sect. \ref{lightyield}). \\
Nevertheless the active fluor concentration, measured by spectrophotometric measurements,
is stable in time, after assessment of the Gd-doped LS, at the level of $\sim 5\%$.

\begin{table}[!ht]
\begin{tabular}{|l|c|c|}
\hline  
                 & UV/VIS    & GC\\   
                 & PPO [g/l] & PPO [g/l]\\
\hline\hline
T40 unloaded & $1.64\pm	0.02$ & $1.51\pm 0.05$\\
T40 October 2005 & $1.07\pm	0.10$ & $0.99 \pm 0.03$\\
T40 January 2006 & $1.07\pm	0.02$ & \\
T40 November 2006& $0.98\pm	0.01$ & \\
T40 January 2007 & $1.00\pm	0.02$ & \\
T40 June 2007	 & $1.00\pm	0.01$ & \\
\hline\hline
T3131 unloaded 	 & $1.83\pm	0.02$ & $1.91 \pm 0.06$\\
T3131 October 2005 &              & $1.72 \pm 0.05$\\
T3131 January 2007&$1.59\pm	0.02$ & \\
T3131 June 2007	&$1.60\pm	0.02$ & \\
\hline
\end{tabular}
  \centering 
  \caption{Spectrophotometric (UV/VIS) and gas-chromatographic (GC) determination  of fluor concentration of samples extracted from tanks T40 and T3131.}\label{measfluorconc}
\end{table}   

\subsection{The light yield}\label{lightyield}
Light output of Gd-doped LS was determined, comparing pulse height amplitude (PHA) spectra collected irradiating Gd-doped and undoped samples in vials of same geometry, with a $^{137}$Cs $\gamma$ source. The LY of Gd-doped LS, compared with the LY of the undoped one, is given by the ratio of the position of the Compton edges in energy spectra.  The PHA spectra of Gd sample and reference were in turn measured in $\approx 10$ cm$^3$ volume cells optically coupled to a Philips XP3462PB PMT. The PMT signal was processed by a standard nuclear spectroscopy electronic chain. The stability of the electronic chain (PMT included) is checked and corrected by cross-calibrating with a 1 inch CsI crystal irradiated by the same source. With the same technique we measured the LY of the undoped LS.
Table \ref{LY_Gd} reports the LY of samples extracted from tank T40 and T3131: the impact of Gd doping on the light yield is the reduction of $10\%$ and $15\%$ respectively. 
It should be noted that in tank T3131 a larger quantity of free carboxilic acid (a quenching agent) has been added to better stabilize the Gd-complex in the LS solution.
\begin{table}[!ht]
\begin{tabular}{|l|c|c|}
\hline  
DATE & LY & REMARK\\
            & [\%] & \\   
\hline\hline	 
12/05/05	& $94\pm10$ & T40 just after doping\\
24/10/05	& $88\pm10$ & T40 sample from extraction\\	
31/01/06	& $90\pm10$ & T40 sample from extraction\\	
14/11/06  & $90\pm10$ & T40 sample from extraction\\ 
\hline\hline
Oct 05    & $85\pm10$& T3131 just after doping\\
Nov 06    & $88\pm10$& T3131 sample from extraction\\
\hline
\end{tabular}
  \centering 
  \caption{Survey of the LY of Gd-doped liquid scintillator from tank T40 and tank T3131. LY values are given relative to the undoped LS yield.}\label{LY_Gd}
\end{table}
From table \ref{LY_Gd} we can conclude that the LY of Gd-LS1 and Gd-LS2 is consistent with a constant behaviour along all the observation period; the average values of the LY, with respect to the undoped samples,  are $90 \pm 5 \%$ and $ 86 \pm 7\%$  respectively.
\section{The \textit{in-tank} measurements}\label{intank}
A second set of mesurement has been performed directly on the two scintillation counters, in order to evaluate the global performances of the detector, in particular its response to neutrons.

\begin{figure}[h]
    \begin{center}
       \vspace{-0cm}
      \includegraphics*[width=7cm,angle=0,clip]{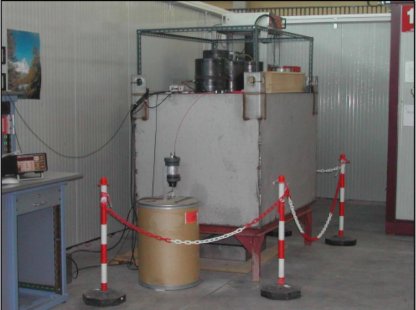}
    \end{center}
    \vspace{-0cm}
    \caption{Tank T40 in the LVD test facility.}
    \label{counter}
\end{figure}

The {\it{in tank}} measurements have been carried out with different PMTs: Photonis XP3550B (118 mm), and FEU (150 mm) and two different electronic set up: a 80 MHz charge-time digitizer\cite{Bigongiari}
and a 300 MHz digital oscilloscope, 500 MS/s.\\
The internal trigger is based on the 3-fold coincidence among the counter's PMTs while three sources of external trigger are used:
\begin{itemize}
\item {\it{the muon trigger}}\\
signals induced by atmospheric muons are used for calibration purposes.
Two plastic scintillators (20 x 20 cm$^\mathrm{2}$ on top and 36 x 36 cm$^\mathrm{2}$ on bottom of the counter) allow to select vertical cosmic ray muons in the counter placed in the external laboratory, T40 (see fig. \ref{counter}).
For T3131, placed inside the LVD array, atmospheric muons are recognized as the coincidence among two or more counters.
\item {\it{the background trigger}}\\
The background is studied by generating spurious triggers at a fixed rate during the measurements.
\item {\it{the fission trigger}}\\
Two $^\mathrm{252}$Cf neutron sources, with very low fission activities ($0.1$ fissions/s for T40 and 0.003 fissions/s for T3131) have been placed inside the two counters through a stainless steel rod. 
$^{252}$Cf decays $\alpha$ in the 97\% of cases and by spontaneous fission in the remaining 3\%.
Each source is surrounded by a Surface Barrier Counter (SBC) which
disentangles the fission's products from the alpha particles generating the trigger \cite{Badino}. 
In the measurement with the oscilloscope, also the signal due to $\gamma$s emitted by the $^\mathrm{252}$Cf source are used for the calibration.
\end{itemize}

\subsection{Neutron detection efficiency} \label{neu_eff}

\begin{figure}[!h]
\begin{center}
\includegraphics*[width=7.cm,angle=0,clip]{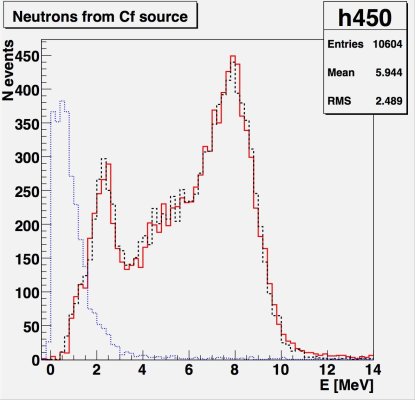}\\
\includegraphics*[width=7.cm,angle=0,clip]{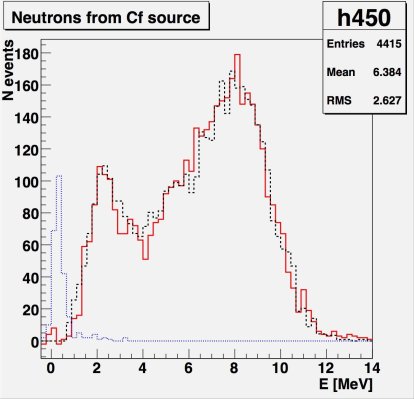}
\end{center}
\caption{ \label{neutron} Typical energy distribution of signals due to neutron captures (red-continuous) after backgound subtraction, detected by T40, on the top, and T3131, on the bottom, compared with the Monte Carlo simulation (black-dashed). 
The background spectra are shown in bleu-dotted.}
\end{figure}

$^{252}$Cf fissions are accompanied by the evaporation of neutrons: their multiplicity distribution is a Gaussian with mean=3.735 and $\sigma$=1.24 and their average energy $\bar E$=2.14 MeV \cite{Hoffman}. 
These neutrons are completely thermalized after about 10 $\mu$s inside the scintillator and, if the the $^{252}$Cf source is placed in the center of the counter, their probability to leave the counter before being captured is $ <1\cdot10^\mathrm{-4}$.\\
Neutrons can be captured by Gd, H, C and finally by the iron of the counter itself, nevertheless captures on C and Fe are largely less probable ($<$ 0.1\%).
Neutron captures on H are followed by the emission of the 2.23 MeV $\gamma$ quantum from the d de-excitation; while captures on Gd (in particular $^\mathrm{155}$Gd and $^\mathrm{157}$Gd) originate $\gamma$ cascades of total energy $\sim$ 8 MeV.\\
In fig. \ref{neutron}
we show an example of the energy spectra due to signals detected within 95 $\mu$s after the fission trigger when the $^\mathrm{252}$Cf sources are placed in the center of the counters. 
The observed spectra, after background subtraction, are compared with the Monte Carlo simulations.
The background spectra, obtained during the same measurements and normalized to the same number of triggers, are also shown.\\
A complete Monte Carlo simulation based on GEANT4 has been developed to interpret the experimental results. 
Neutron capture processes, due to Gd, H, Fe and C are completely described as well as the light collection, PMTs and front-end electronics \cite{Amanda}.

\begin{figure}[!h]
\begin{center}
\includegraphics*[width=8.cm,angle=0,clip]{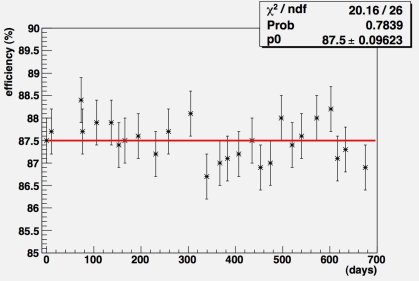}\\
\includegraphics*[width=8.cm,angle=0,clip]{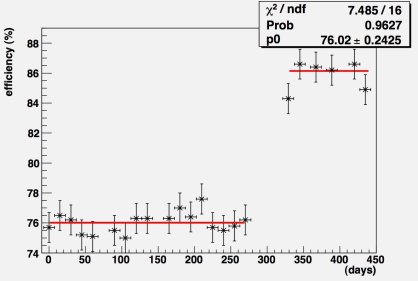}
\end{center}
\caption{ \label{efficiency}Neutron detection efficiency, $\eta_\mathrm{n}$, survey for T40 (top) with $^\mathrm{252}$Cf source placed in the center and the first point corresponding to August 1$^\mathrm{st}$ 2005 and for T3131 (bottom) with $^\mathrm{252}$Cf source placed in contact with a wall during the first 270 days of measurement, then moved to the center of the counter, and the first point corresponding to June 28$^\mathrm{th}$ 2006.}
\end{figure}

The data shown in fig. \ref{neutron} regard measurements carried out with the digital oscilloscope; they correspond to 3000 fissions in T40 and 1400 in T3131. 
These data have been collected during about 2 days in T40 and almost one week in T3131 because of the different activities of the two $^\mathrm{252}$Cf sources and of the oscilloscope dead time.
The two components, capture on H (2.2 MeV) and capture on Gd ($\sim$ 8 MeV), are clearly distinguishable, together with the different impacts of background in the two experimental conditions.
Moreover, by studying the width of the 2.2 MeV gamma peak, one can observe that T40 has a better energy resolution than T3131. 
This is due to the different PMTs used in the two measurements, Photonis XP3550B in T40 and FEU 125 in T3131, that have different light collections.\\
We measure the neutron detection efficiency , $\eta_\mathrm{n}$, of the counter by comparing the detected multiplicity distribution (inside defined time windows started by the fission trigger) with the one expected by the $^\mathrm{252}$Cf source.
The long term stability of $\eta_\mathrm{n}$ of the two counters has been studied during about two years. The result is shown in fig. \ref{efficiency}.\\
T40 has been doped in May 2005 and monitored since August 2005 with the source always placed in its center. On June 13$^{th}$ 2006 (day 317 in the figure) the three PMTs FEU 49b were substituted by three Photonis XP3550B.
\\
T3131 has been doped in October 2005; in June 2006 the $^\mathrm{252}$Cf source has been erroneously inserted in a lateral position inside the counter, in contact with one wall. Due to the different position of the source, the neutron detection efficiency results sensibly lower in the first period (from 76.0$\pm$0.2 to 86.$\pm$0.5).
Nevertheless that affects the absolute value of $\eta_\mathrm{n}$ and mean capture time, but it has no impact on the quality of the stability monitor.
On April 2007 (day 270 in the figure), the source has been finally moved to the center of the counter.

The results of these measurements, for different electronics set ups and for the sources placed in the center of the counters, are summarized in table \ref{results}.
The difference between the efficiencies measured by the two electronics in the same counter, is due to the different dead time: 250 ns (integration time) for the oscilloscope and 1.0 $\mu$s for the charge-time digitizer. 
If we force, in the analysis of the oscilloscope data, the same dead time of the charge-time digitizer, the two values coincide\footnote{If we take into account that neutrons can be detected only during a time window of 95 $\mu$s after the occurrence of each fission, the unbiased efficiencies become 94.3 for T40 and 91.7 for T3131.}.\\
The difference between the efficiency measured in the two counters is probably due to the combination of the higher light yield of the liquid scintillator and the higher light collection of PMTs of T40.
In both counters the long term stability survey does not show any indication of degradation of the n-capture efficiency.\\ 
This result is highly relevant because the neutron detection efficiency is of utmost importance for $\bar\nu_\mathrm{e}$ detectors, even if it does not represent a strong indication of the stability of the optical characteristics of the liquid scintillator; indeed the efficiency remains almost constant even reducing the light collection of a factor of two.
\begin{table}[!h]
\begin{center}
\begin{tabular}{|l|c|c|c|}
\hline 
 & \multicolumn{3}{c|}{neutron detection efficiency} \\
\hline 
 counter & charge digitizer & oscilloscope &\\
               &$\eta_\mathrm{n}~[\%]$  & $\eta_\mathrm{n}~[\%]$ &\\  
               \hline
 T40 & 87.5 $\pm$ 0.1 & 92.0 $\pm$ 0.7 &\\ 
 \hline
 T3131& 86.1 $\pm$0.5& 88.7 $\pm$0.9&\\ 
\hline\hline
& \multicolumn{3}{c|}{neutron capture time} \\
\hline
counter & charge digitizer& oscilloscope& simulations\\
& $\tau_\mathrm{n}~[\mu s]$ & $\tau_\mathrm{n}~[\mu s]$&$\tau_\mathrm{n}~[\mu s]$\\ 
\hline\hline
T40 & 24.5 $\pm$ 0.1 & 25 $\pm$1&24.6 $\pm$0.5 \\
\hline
T3131& 27.0 $\pm$ 0.2 & 27.6 $\pm$0.8&27.9 $\pm$0.6 \\
\hline
\end{tabular}
\end{center}
\caption{Neutron detection performances of Gd-doped LVD counters, T40 and T3131. The $^\mathrm{252}$Cf sources were placed in the center of each counter.}
\label{results}
\end{table}
\subsection{Mean capture time} \label{neu_captime}
\begin{figure}[!h]
\begin{center}
\includegraphics*[width=7.2cm,angle=0,clip]{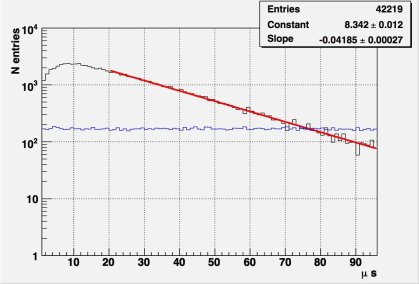}\\
\includegraphics*[width=7.2cm,angle=0,clip]{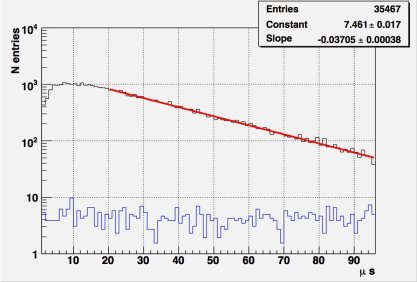}
\end{center}
\caption{ \label{time}Time distribution of the intervals between n-capture events and relative fission for T40, top, and T3131, bottom, for about 15000 fission.
The distributions of pure background events, normalized to the same number of triggers and subtracted, are also shown.}
\end{figure}

The mean n-capture time is obtained by the fit of the distribution of time difference between the n-capture and its relative fission, measured over about 15000 fissions, and taking into account that the first $\sim$10 $\mu$s are spent by $^\mathrm{252}$Cf's neutrons to reach complete thermalization.
As an example two of these measurements, one for T40 and the other for T3131, are shown in fig. \ref{time}, compared with the corresponding background distributions. 

\begin{figure}[!h]
\begin{center}
\includegraphics*[width=7.2cm,angle=0,clip]{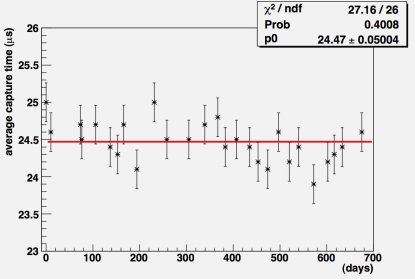}\\
\includegraphics*[width=7.2cm,angle=0,clip]{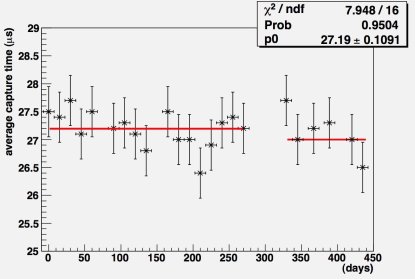}
\end{center}
\caption{ \label{tau}Mean capture time, $\tau_\mathrm{n}$, survey for T40 (with $^\mathrm{252}$Cf source placed in the center and the first point corresponding to August 1$^\mathrm{st}$ 2005) and for T3131 (with $^\mathrm{252}$Cf source placed in contact with a wall during the first 270 days of measurement, then moved to the center of the counter, and the first point corresponding to June 28$^\mathrm{th}$ 2006.}
\end{figure}

The results of these measurements, performed during almost two years on the two counters are shown in fig. \ref{tau} and summarized in table \ref{results}.
The mean capture time, $<\tau_\mathrm{n}>$, averaged over the entire period, is 24.5 $\pm$ 0.1 $\mu$s for T40 and 27.0 $\pm$ 0.2 $\mu$s for T3131, when the $^\mathrm{252}$Cf source is placed in the center of the counters.

This difference is in perfect agreement with Monte Carlo expectations (see table \ref{results}) and it is due to the different Gd concentration in the liquid scintillator of the two counters.\\
The value of $\tau_\mathrm{n}$, that is measured with very good precision, strongly depends on the Gd concentration in the liquid and is almost independent on variation of the gain of scintillator, PMTs and electronics. 
For these reasons the long term survey of $\tau_\mathrm{n}$ states the stability of Gd concentration in solution and it confirms the results obtained by titrations reported in sect. \ref{fluorconc}.
\section{Conclusions} \label{conclusion}
Two large samples ($1.2$ ton each) of the two liquid scintillators used in the LVD experiment have been doped with Gd at the $0.1$\% level. They are contained in two counters of the type in use in LVD. The first one (T40) has been doped in May 2005 and has been placed in an experimental hall in the external lab, the second one (T3131) has been doped in Oct 2005 and has been placed in Hall A of the underground lab. \\
They have been continuously monitored for a total time of about $700$ days.\\
We measured the chemical, optical and fluorescent properties of the LS periodically extracting small samples from the counters. The study of the Gd concentration through EDTA titration does not show any variation. The light transmittance (T), measured with a $10$ cm optical path, shows for counter T40 an initial  small decrease with respect to the undoped LS, but then the subsequent measurements are consistent with a constant behaviour (C.L. 99\% at the reference wavelength of 425 nm); for counter T3131 the stability of T is confirmed at 96\% C.L.. 
An initial decrease and subsequent stability is observed also studying the primary fluor concentration, measured by two independent techniques (spectrophotometry and gas-chromatography); this effect needs further investigation. The Gd doping has a little impact on the LS light yield, measured observing the response to a $^{137}$Cs $\gamma$ source; we found $LY_{doped}/LY_{undoped}=90\%$ and $85\%$ respectively for target T40 and T3131 respectively. \\
Together with these laboratory measurements we also studied the global performances of the two scintillation counters. We used a $^\mathrm{252}$Cf neutron source and we periodically measured the mean neutron capture time and neutron detection efficiency. They both do not show any hint of variation and degradation. It is important to point out that tank T40 was operated at surface laboratory in a building where the environmental temperature ranges from 18 to 32 $^\circ$C from winter to summer. 

With this work we demonstrate that the adopted Gd-doping tecnique is robust, reproducible and it can be used to dope very large quantities of liquid scintillator (hundreds of tons). 
This is the largest sample of Gd-doped liquid scintillator, stable for 2 years, ever produced and continuously operated.

\section{Acknowledgments} \label{ackno}
This work has been supported by INFN, INAF-IFSI-Torino and INR.
We wish to thank the LNGS Chemical Service for the continuous support, dr. Corrado Salvo and dr. Galia Novikova, for their help and productive discussion, Dr. Nina Nesterova for the distillation of carboxylic acid, and Antonio Chiarini and Alberto Romero for their valuable technical assistance. We wish to thank also Dr. Venja Berezinsky and prof. Alessandro Bettini for their partecipation in early phases of the organization of INFN/INR group. \\


\end{document}